\documentclass[journal=jacsat,manuscript=article]{achemso}

\usepackage[T1]{fontenc}       
\usepackage[usenames, dvipsnames]{color}

\author{Hrishit Banerjee}
\affiliation{Department of Condensed Matter Physics and Material Sciences, S N Bose National Centre for Basic Sciences}
\author{Sudip Chakraborty}
\affiliation{Materials Theory Division, Department of Physics and Astronomy, Uppsala University, Box 516, 75120 Uppsala, Sweden}
\author{Tanusri Saha-Dasgupta}
\affiliation{Department of Condensed Matter Physics and Material Sciences, S N Bose National Centre for Basic Sciences}
\email{t.sahadasgupta@gmail.com}

\title[ \textsf{achemso} demo]
{Cationic Effect on Pressure driven Spin-State Transition and cooperativity in Hybrid Perosvkites}

\abbreviations{IR,NMR,UV}

\begin{document}


\begin{abstract}
Hybrid or metal organic framework (MOF) perovskites of general composition, ABX$_3$, are known 
to show interesting properties which can lead to variety of technological applications. Our first 
principles study shows they are also potential candidates for exhibiting cooperative spin-state 
transitions upon application of external stimuli. We demonstrate this by considering two specific 
Fe-based MOF perovskites, namely Dimethylammonium Iron Formate, [CH$_{3}$NH$_{2}$CH$_{3}$][Fe(HCOO)$_3$] 
and Hydroxylammonium Iron Formate, [NH$_{3}$OH][Fe(HCOO)$_3$]. 
Both the compounds are found to undergo high-spin (S=2) to low-spin (S=0) transition at Fe(II) 
site upon application of moderate strength of hydrostatic pressure, along with large hysteresis. 
This spin-state transition is signaled by the changes in electronic, magnetic and optical properties.
We find both the transition pressure and the width of the hysteresis to be strongly dependent
of the choice of A-site cation, Dimethylammonium or Hydroxylammonium, implying tuning of 
spin-switching properties achievable by chemical variation of the amine cation in the structure. 
Our finding opens up novel functionalities in this family of materials of recent interest, which 
can have important usage in sensors and memory devices.
\end{abstract}

\section{Introduction}
Metal organic framework compounds, built from inorganic and organic components, form an active area of research which has undergone a rapid growth in last decade.\cite{Cheetham-ChemSci,Shekhah} Bulk of the 
study in this field are focused on open systems with large porous region having potential applications in gas storage,\cite{Li} chemical 
sensing,\cite{Kreno} catalysis \cite{Ranochhiari} etc. In recent time attention has been given also to dense hybrid frameworks with limited 
porosity. Some of these dense hybrid compounds adopt the celebrated perovskite geometry of general formula ABX$_3$.\cite{Tejuka} This has 
opened up an emerging research area on hybrid perovskites, parallel to the well-established field of inorganic perovskite oxides. The halide 
hybrid perovskites of general composition, [AmH]MX$_3$ (AmH$^{+}$ = protonated amine, M = Sn$^{2+}$ or Pb$^{2}$, and X = Cl$^{-}$ or Br$^{-}$ 
or I$^{-}$) have already attracted great deal of attention due to their potential use in solar 
cell applications.\cite{Yin} The other class of hybrid perovskites discussed in very recent time due to their attractive ferroic
properties, are transition metal formates [AmH]M(HCOO)$_3$ (M = Mn, Cu, Ni, Fe, Co), where MO$_6$ octahedra are linked via formate bridges, 
and the protonated amine molecules sit in the hollow formed by the octahedral framework, establishing hydrogen bonding with formates.\cite{Jain}
The presence of organic components in the structure offers greater structural flexibility, and thus better tunability of properties by external
means, as compared to that of inorganic perovskites. The flexibility of hybrid perovskite to undergo large structural changes in response to
external stimuli has been already reported.\cite{Cohen} Furthermore, it is possible to tailor properties by changing the amine molecule, thereby
changing the strength and cross-linking of hydrogen bondings in the structure, yet maintaining 
the basic topology.\cite{Cheetham-JACS} Needless to say, much remains to be explored
in terms of functionalities that can be achieved in this new class of perovskite materials. 

In this study, we focus on one such unexplored area, namely the external stimuli driven spin-crossover (SCO).
The tremendous flexibility of the organic linkers makes the transition metal based formate hybrid 
perovskites ideally suited for triggering spin-crossover from a high-spin (HS) to low-spin (LS) 
state at the transition metal site by external perturbation. A much discussed aspect in this context 
is the issue of cooperativity in SCO phenomena.\cite{Banerjee} The cooperativity in SCO phenomena 
makes it a spin transition rather than spin-crossover, which may show up with associated hysteresis, 
having important implications in designing memory devices.\cite{Kahn} For device
applications, the challenge is to make the width of the hysteresis large for obtaining the memory effect
over a wide range of external stimuli as well as transition to occur at a value of the 
stimuli that is readily achievable.  In this respect, commonly
explored candidates are SCO polymers or 3D coordination compounds which are expected to provide 
better connectivity compared to molecular crystals with isolated molecular units.\cite{Martinez} 
Considering the dense topology of the newly discussed hybrid perovskites, they can be a potential 
alternative to SCO polymers or 3D coordination compounds in exhibiting cooperative spin-state 
transitions. This will add novel functionality in this novel and interesting class of compounds. 

In light of the above, we venture on the study of SCO properties of transition metal based formate hybrid 
perovskites, through first-principles calculations. In particular we consider Fe$^{2+}$ based hybrid 
perovskites. The choice is prompted by the fact that SCO transitions in Fe(II)-based compounds, 
having 6 $d$-electrons in Fe(II) ion, and showing a transition from a LS (S = 0) to a HS (S = 2) 
state are pronounced and abrupt, making them suitable for applications. In our study, we consider 
hydrostatic pressure as an external stimuli. To the best of our knowledge, no study exists so far on 
the pressure effect on hybrid perovskites, although hydrostatic pressure is considered as one of the 
effective means to tune properties in inorganic perovskites.\cite{Manoun} Additionally, to study the 
influence of changing the primary organic component, namely the amine cation, located in the perovskite 
cavity, we consider two different cations,  Dimethyl-ammonium (CH$_{3}$NH$_{2}$CH$_{3}$), and 
Hydroxylamine (NH$_{3}$OH). It is worth to note that our conscious choice of these two different molecular cations, 
leads to a change in the tolerance factor of the Fe formate perovskite structures, as the effective molecular 
radii for Dimethyl-ammonium and Hydroxylamine are different, being 272 pm and 215 pm, respectively. As was 
pointed by Kieslich and co-workers \cite{Cheetham-ChemSci}, change in mechanical properties, particularly 
in rigidity can be achieved by changing cations of different effective molecular radius. 

Our density functional theory (DFT) based computational study that takes into account all structural 
and chemical aspects in full rigor, shows that pressure induced spin-state transitions are achieved in 
both Dimethyl-ammonium Iron Formate (DMAFeF) and Hydroxylamine Iron Formate (HAFeF) compounds 
for modest critical pressure range of 2-7 GPa, associated with large hysteresis of 2-5 GPa. The latter 
implies that these compounds should exhibit spin-switchablility over a wide range of operating 
pressure.  Our findings highlight the possible technical use of spin-switching functionalities 
of hybrid perovskite compounds with accompanied changes in electronic, magnetic and optical 
properties, in sensors and memory devices. Interestingly, the flexibility in choice of the A 
site cation, {\it i.e.} the protonated amine molecule adds another dimension, namely the tuning and 
modulation of spin-switching properties.  

\section{Computational Methodology}

Our first-principles calculations were carried out in the plane wave basis as implemented in the Vienna Ab-initio Simulation Package (VASP)\cite{kresse1,kresse2} with projector-augmented wave (PAW)\cite{blochl} potential. The exchange-correlation functional was chosen to be that of generalized gradient approximation (GGA) implemented following the Perdew-Burke-Ernzerhof \cite{pbe} prescription. For ionic 
relaxations, internal positions of the atoms were allowed to relax until the forces became less than 0.005 eV/A$^0$. Energy cutoff of 500 eV, and $4\times 4\times 2$ Monkhorst-Pack k-points mesh were found to provide a good convergence of the total energy in self-consistent field calculations. To take into account of the correlation effect at Fe sites beyond GGA, which turn out to be crucial for the correct description of the electronic and magnetic properties, calculations with supplemented Hubbard U (GGA + U) a la Liechtenstein et al \cite{liechtenstein} were carried out, with the choice of U = 4 eV and Hund's coupling parameter J$_H$ = 1 eV. In order to study the effect of hydrostatic pressure, calculations were done by first changing the volume of the unit cell isotropically and then relaxing the shape of the cell together with the ionic positions. The estimate of applied hydrostatic pressure for each compressed volume was 
obtained from the knowledge of the calculated bulk modulus. The bulk modulus was calculated by varying the volume of the unit cell 
and relaxing the ionic positions at each volume. Accurate self-consistent-field calculations were carried out to obtain the total energy of the systems at each volume. The energy versus volume data was fitted to the third order Birch-Murnaghan isothermal 
equation of state\cite{murna}, given by,
$$ E(V)=E_0+\frac{9V_0B_0}{16}\lbrace [(\frac{V_0}{V})^{2/3}-1]^3 B'_0 + [(\frac{V_0}{V})^{2/3}-1]^2 [6-4(\frac{V_0}{V})^{2/3}] \rbrace $$
where {\it V$_{0}$} is the equilibrium volume, {\it B$_{0}$} is
the bulk modulus and is given by $B_{0}=-V(\delta P/\delta V)_{T}$
evaluated at volume {\it V$_{0}$}. {\it B$^{'}_{0}$} is the
pressure derivative of {\it B$_{0}$} also evaluated at volume {\it
V$_{0}$}.

\section{Results and Discussions}

{\bf Crystal Structure of DMAFeF and HAFeF-}
As essential pre-requisite for the first-principles study is the accurate information of the crystal structure. An interesting feature 
of [AmH]M(HCOO)$_3$ compounds is the order-disorder transition of the A-site amine cations through ordering of hydrogen bonds.\cite{Jain} 
While the crystal structure data for disordered phase of  DMAFeF\cite{Jain} is available, no such data exists for the corresponding ordered phase. Moreover, in case of HAFeF, no crystal structure data has been reported till date. 
Therefore we started with crystal structure data for the ordered phases of 
DMAMnF\cite{Sanchez} and HAMnF\cite{Gao}, the Mn analogues of DMAFeF and HAFeF. We relaxed the structure completely after replacing Mn atoms with 
Fe atoms, which gave the first-principles predicted ordered structures of DMAFeF and HAFeF. Mn being next to Fe in the periodic table, this forms a 
legitimate approach. We carried out a complete structural relaxation, which involved
relaxation of the unit cell volume and shape, as well as atomic positions. We found that though the symmetries do not change between
Fe compounds and their corresponding Mn counterparts, there is appreciable change in the volume of the unit cells, as expected. 
DMAFeF and HAFeF crystallize in two different monoclinic space groups, DMAFeF being in Cc space group and HAFeF being in P2$_{1}$ space group. 
The lattice constants for DMAFeF are found to be, a=14.464A$^0$, b=8.355A$^0$, c=8.975A$^0$, with the angle 
$\gamma=119.8^0$, whereas for HAFeF the lattice constants are found to be 
a=7.812A$^0$, b=7.961A$^0$, c=13.173A$^0$, with angles $\alpha=\beta=\gamma=90^0$. The calculated 
crystal structures as .cif files can be found in the supplementary information (SI).

\begin{figure}
\begin{center}
\includegraphics[width=0.9\linewidth]{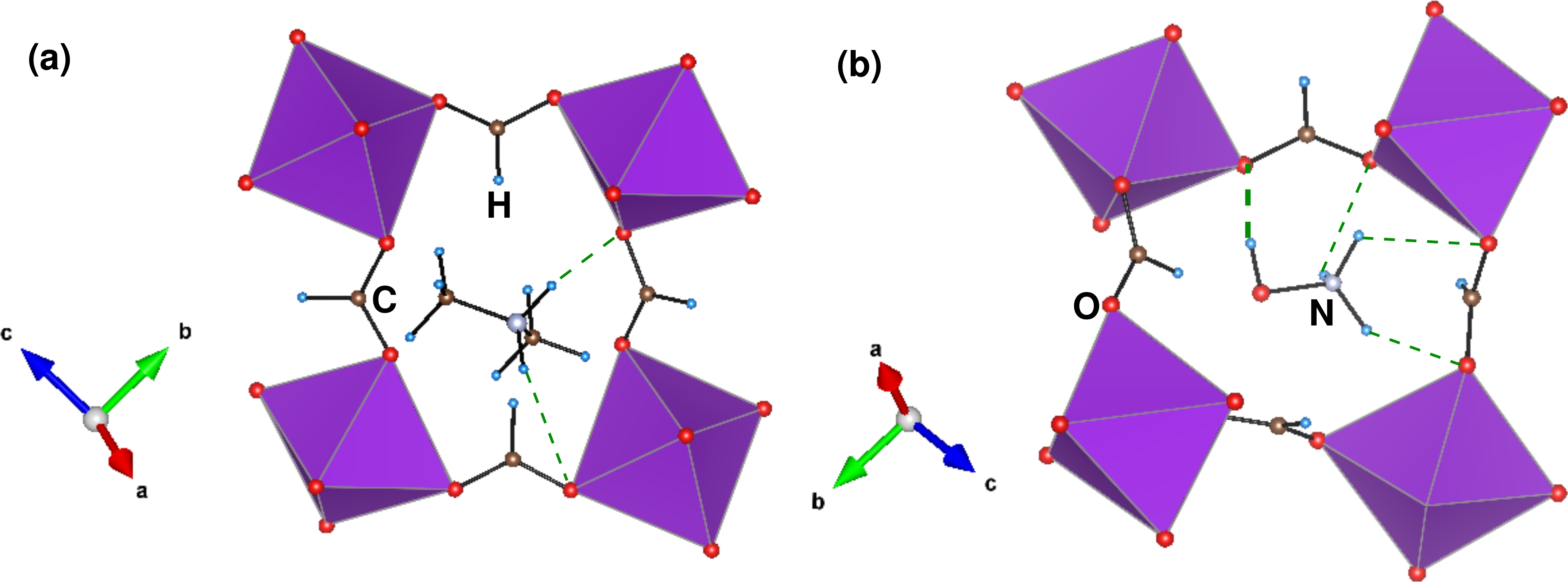}
\caption{Computed crystal structures of DMAFeF [panel (a)] and HAFeF [panel (b)] in A-site ordered phase. The FeO$_6$ octahedra are 
connected to each other by the formate ligands while the DMA or HA cation sit in the hollow formed by the octahedra. Various atoms 
have been marked. N-H$\cdot\cdot\cdot$O and O-H$\cdot\cdot\cdot$O bonds are represented by dashed lines, with thickness of lines 
indicating the strength of the bonds.}
\end{center}
\end{figure}

As shown in Fig. 1, in both the framework compounds each FeO$_6$ octahedra is connected to 
neighboring FeO$_6$ octahedra via HCOO$^{−}$ ligand bridges. This forms a three-dimensional 
ReO$_3$-type network, with Dimethyl-ammonium or Hydroxylamine cations occupying the centers of the ReO$_3$-type cavities.
In DMAFeF, two bridging N-H$\cdot\cdot\cdot$O hydrogen bonds from each DMA cation are formed, while in HAFeF, three N-H$\cdot\cdot\cdot$O hydrogen bonds,
and a O-H$\cdot\cdot\cdot$O hydrogen bond are formed from each HA cation. The nature of hydrogen bondings is expected to be different
between N-H$\cdot\cdot\cdot$O and O-H$\cdot\cdot\cdot$O due to the less polar nature of N-H bond as compared to O-H bond.
Thus the O-H$\cdot\cdot\cdot$O hydrogen bond is stronger than N-H$\cdot\cdot\cdot$O hydrogen bond. 
The lattice for HAFeF is therefore expected to be more rigid compared to the lattice for DMAFeF, having 
important bearing on SCO phenomena as we will discuss in the following.

\begin{figure}
\begin{center}
\includegraphics[width=0.6\columnwidth]{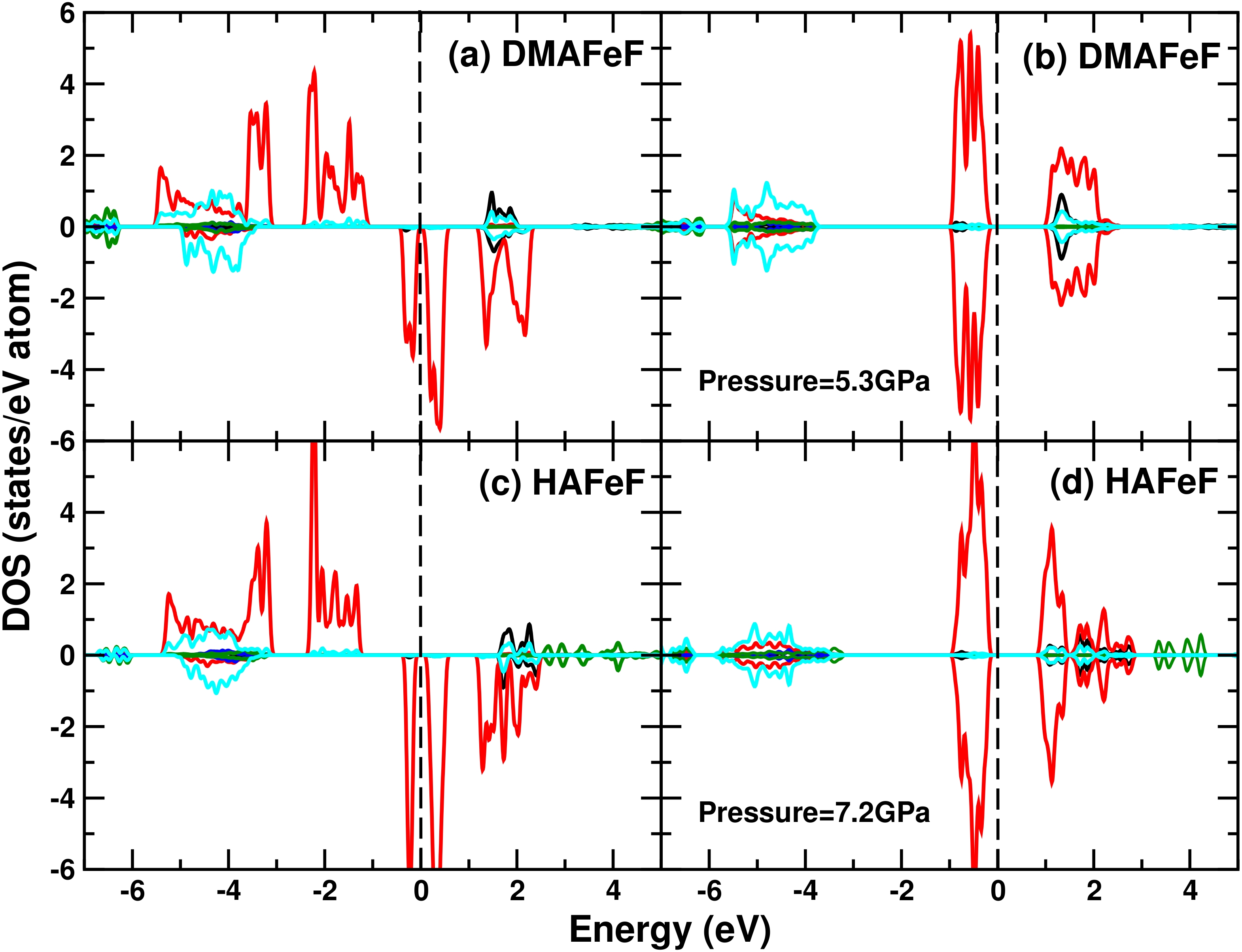}
\caption{Projected density of states of DMAFeF and HAFeF at ambient and high pressure conditions. Panels (a) and (c) are for 
ambient pressure, while panels (b) and (d) are for high pressure. The DOS projected to Fe $d$, O $p$, C $p$, N $p$ and H $s$
are marked in in red, cyan, black, green and blue, respectively. The zero of the energy is set at Fermi energy.}
\end{center}
\end{figure}
\begin{figure}
\begin{center}
\includegraphics[width=0.7\linewidth]{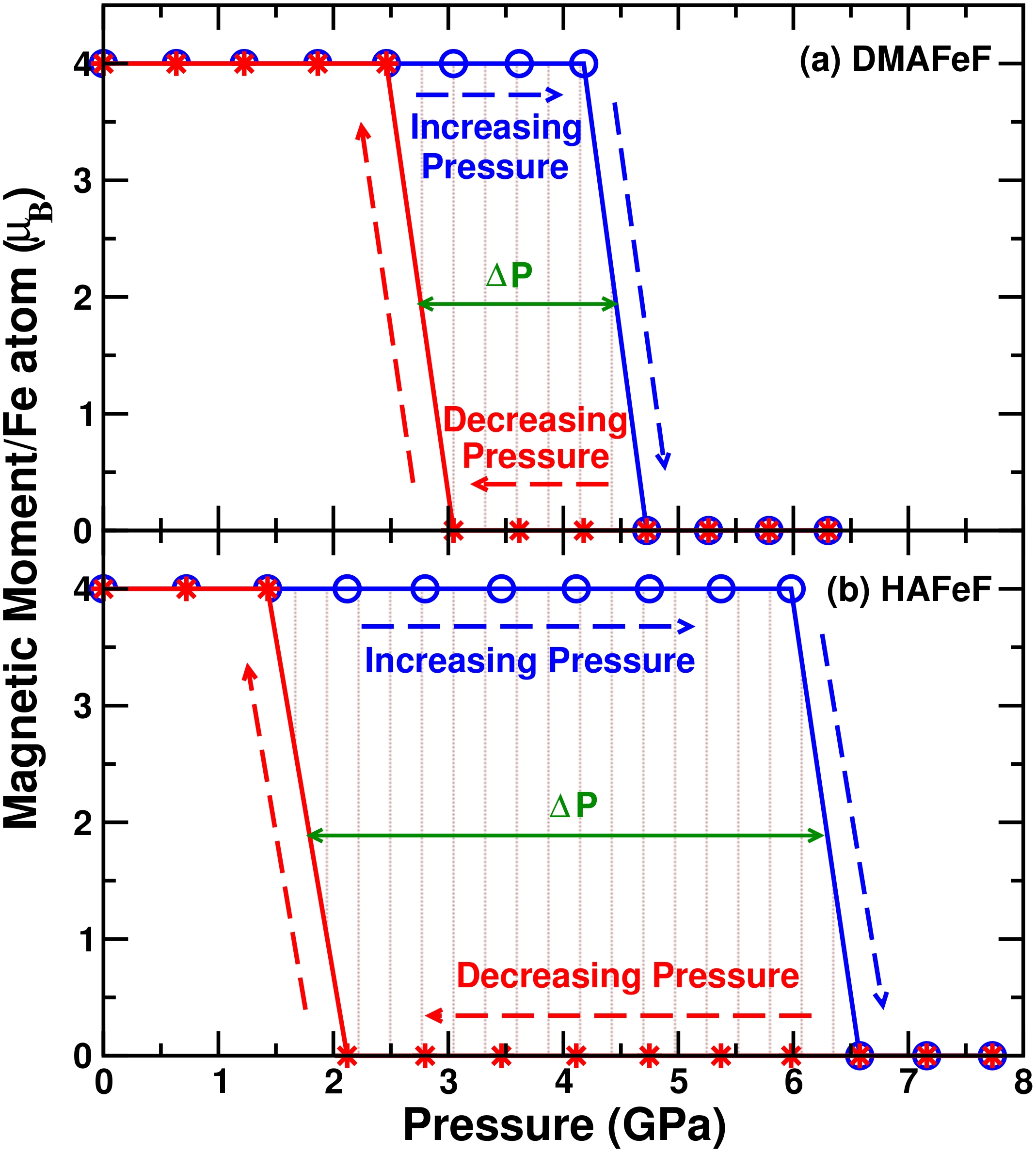}
\caption{Computed magnetic moment at the Fe site plotted as a function of pressure for DMAFeF [panel (a)] and HAFeF [panel (b)]. 
The data are plotted for two different paths. The data points in blue
denote the path following increasing pressure, starting from the HS state and the data points in red denote 
the path following decreasing pressure, starting from the LS state. HS $\rightarrow$ LS transitions in both compounds 
exhibit interesting hysteresis effects.}
\end{center}
\end{figure}

{\bf Spin-state transition under pressure and Cooperativity-}
In order to determine the spin-states of the Fe atoms in DMAFeF and HAFeF, we calculated the spin-polarized electronic
structures. The spin-polarized density of states at ambient pressure condition for DMAFeF and HAFeF are shown in 
panels (a) and (c) of Fig. 2, respectively. The states close to Fermi level (E$_F$) are dominantly of Fe $d$ character,
which are found to be strongly spin-polarized. The octahedral coordination of oxygens atoms around Fe, groups the Fe $d$ states
into states of $e_g$ and $t_{2g}$ symmetries. The Fe $d$ states are found to be completely occupied in the majority spin
channel, with empty Fe $e_g$ states, and partially filled Fe $t_{2g}$ states in the minority spin channel. The distortion in
the FeO$_6$ octahedra causes further splitting within states of Fe $t_{2g}$, leading to a small gap at the Fermi region in the minority spin channel. The insulating solution is obtained for both compounds at ambient condition, with a large band gap ($\approx$ 2 eV) 
between occupied Fe $d$ and empty C $p$ states in majority spin channel and a tiny band gap ($\approx$ 0.1 eV)
within the Fe $t_{2g}$ states in the minority spin channel. This suggests at ambient condition spin-state of Fe in both DMAFeF and
HAFeF to be HS. The calculated total magnetic moment (M) turned out to be 4 $\mu_B$ per Fe atom, for both
the compounds, in conformity with the stabilization of HS (S=2) state of Fe. Panels (b) and (d) of Fig. 2 show the 
spin-polarized density of states for DMAFeF and HAFeF at high pressure condition. We find the exerting pressure (P) in the 
range of $\approx$ 5-7 GPa causes drastic change in the electronic structure. First of all, for both the compounds the ground states
turned out to be non spin-polarized, with calculated magnetic moments of 0 $\mu_B$. This confirms a spin-state
transition from HS (S=2) state to LS (S=0) state, obtained by application of pressure. In the high-pressure LS state a large
band gap of $\approx$ 1 eV opens up between the fully occupied Fe $t_{2g}$ states, and completely empty Fe $e_g$ states.
The spin-state transition thus should be accompanied by a significant change in the over all band gap, which should be manifested
in corresponding change in optical response. 

In the next step, in order to find out the critical pressure where such spin-state transition happens for the two compounds, 
we increased the pressure in step of 0.6-0.7 GPa, starting from the ambient pressure condition. As shown in the plot of the magnetic
moment (M) versus pressure (P) in Fig. 3, we find a spin-state from HS with total magnetic moment of  4 $\mu_B$/Fe to LS with
a total moment of 0 $\mu_B$/Fe at pressure (P$_{c}\uparrow$) of 4.7 GPa for DMAFeF and 6.6 GPa for HAFeF. This implies a strong influence 
of choice of A cation on the optimal pressure needed for spin-state transition. We then decreased the pressure starting from the highest 
applied pressure. Interestingly we find the optimal pressure required for the transition from LS with a total moment of 0 $\mu_B$/Fe to 
HS with total magnetic moment of 4 $\mu_B$/Fe, happens at a different pressure (P$_{c}\downarrow$) compared to P$_{c}\uparrow$, having 
values 2.5 GPa for DMAFeF and 1.4 GPa for HAFeF. There is reflected as significant hysteresis effect in M-P data in case of both compounds, 
with width of hysteresis being 2.2 GPa for DMAFeFe as compared to 5.2 GPa for HAFeF, the former being more than a factor of 2 smaller
than the latter. Therefore, the choice of A cation has a significant influence on spin-switching properties. 
This constitutes the key finding of our investigation. We note that both P$_{c}\uparrow$ and P$_{c}\downarrow$ are of moderate values for both 
compounds, that can be generated in a laboratory set-up.


\begin{figure}
\begin{center}
\includegraphics[width=0.8\linewidth]{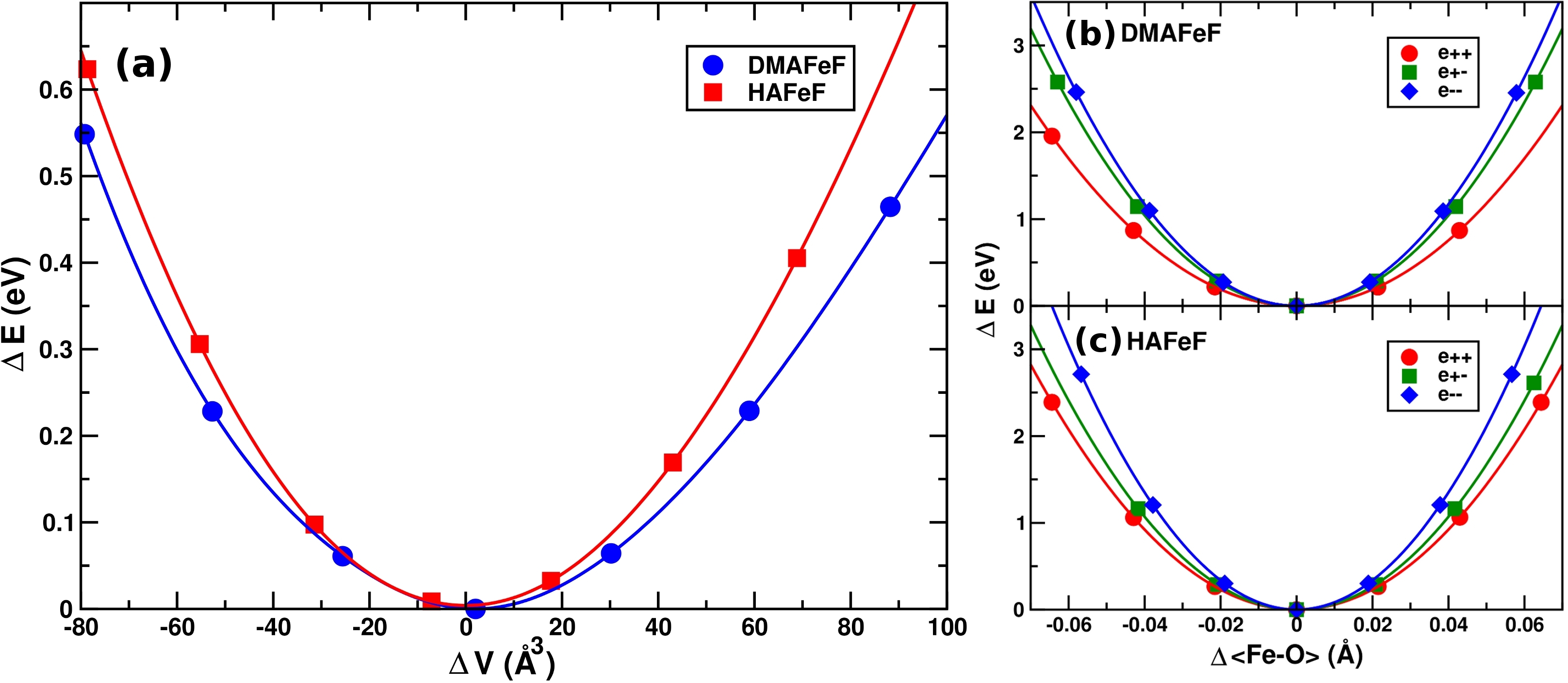}
\caption{Panel (a): Total energy versus volume of the unit cell for DMAFeF and HAFeF. Energy and volumes have been measured with respect to the equilibrium values, $E_0$ and $V_0$, respectively. The symbols denote the DFT calculated data points and the lines depict the that obtained
from fitting with Murnaghan\cite{murna} equation of state. Panel (b): The variation of total energy as a function of variation
of Fe-O bondlength about the equilibrium value, for different spin-state configurations of neighboring Fe atoms, 
LS-LS (diamond), LS-HS (square) and HS-HS (circle). Solid lines are fit to the data points. Panel (c): same as in panel (b), but for HAFeF.}
\end{center}
\end{figure}

{\bf Microscopic Understanding-} In order to understand the microscopic origin of the quantitative differences in response 
of the two formate frameworks considered in this study to the applied pressure, we firstly calculated the mechanical strengths of the 
two compounds. As mentioned previously due to the differential nature of H bonding, the lattice for HAFeF is expected to be 
more rigid compared to that of DMAFeF. This is confirmed by the magnitudes of the calculated bulk moduli of 
the two systems. The fit of the DFT total energy versus volume data to the Birch-Murnaghan equation of 
state,\cite{murna} as shown in Fig. 4(a), gave the bulk modulus to be 21.55 GPa for DMAFeF, and 24.27 GPa 
for HAFeF, with {\it B$^{'}_{0}$} = 5.39 and 1.84, respectively. This would in turn imply that the critical pressure 
needed to cause spin-switching through change in Fe-O bondlength to be larger in the significantly more rigid lattice 
of HAFeF compared to that in comparatively less rigid lattice of DMAFeF, as reflected in different values of P$_{c}\uparrow$ in
two compounds. 
\paragraph{}
Further to elucidate the mechanism by which the cooperativity, manifested in form of hysteresis in the M-P plot,
develops in these materials we computed the elastic and magnetic exchange interactions. The most prevalent idea
in this context attributes the microscopic origin of cooperativity to the elastic interaction between local distortions at 
the SCO centers.\cite{Boukheddaden} However in a very recent work, the importance of magnetic super-exchanges in driving 
cooperativity was uncovered.\cite{Banerjee} Depending on the sign of the spin-dependent elastic interaction, which 
is dictated by the nature of the spin-phonon coupling in the material, the magnetic interaction was found to be influence
the hysteresis in a quantitative or qualitative way. 
As was found in Ref.\cite{Banerjee} the interplay between elastic and magnetic interaction in building up
cooperativity, crucially relies on the spin-dependent rigidity of the lattice. Depending the spin-state of the
spin-crossover ion, which is Fe in the present case, the elastic interaction between two neighboring Fe ions can be different, 
which are labeled as $e_{++}$ ($e_{--}$) for both neighboring sites in HS (LS) state, and $e_{-+}$ for one site in LS 
and another in HS. The size change of the SCO unit upon change of spin-state makes $e_{--}$ > $e_{++}$. It is thus
the value of $e_{-+}$ which decides the nature (sign) of the effective elastic interaction. Following the work
by K Boukheddaden et al.,\cite{Boukheddaden} the effective elastic interaction between two neighboring SCO sites 
is given by, $$K=\frac{1}{8}ln(\frac{e_{+-}^2}{e_{++}\times e_{--}})$$
Thus the effective elastic interaction, $K$ turns out to be of ferroelastic nature for $e_{+-} > \sqrt{e_{++} \times e_{--}}$
and of antiferroelastic nature for $e_{+-} < \sqrt{e_{++} \times e_{--}}$. It was demonstrated\cite{Banerjee} that for 
ferro type elastic interaction, the magnetic interaction becomes operative only in qualitative manner, in terms of enhancing 
the hysteresis width, while for antiferro nature of elastic interaction, the magnetic
exchange is the sole driving force in setting up the hysteresis.
 
In order to have a microscopic understanding of the observed hysteresis in M-P plot of the studied formate frameworks and its
dependence on the choice of the amine cation, we thus first calculated the spin-state dependent elastic interactions for
the two compounds. To do so, we adopted the same procedure as in Ref.\cite{Banerjee}. Electronic structure of the optimized 
crystal structure data shows 
that the LS state is obtained at a high pressure phase for an average Fe-O bond length of 1.9 A$^0$  or less, while the HS state 
is obtained at an ambient pressure phase for for an average Fe-O bond length of 2.1 A$^0$ or more. Keeping this in mind, we constructed
crystal structures setting the average Fe-O bond length at 1.9 and 2.1 A$^0$ to emulate the LS and HS states of neighboring Fe-O$_6$ octahedra, 
respectively. To simulate the LS-HS situation, structure with alternating arrangements of Fe-O$_6$ octahedra having average Fe-O bond lengths of 1.9 and 2.2 A$^0$ 
was constructed. Considering the three model structures with HS-HS, LS-LS and LS-HS arrangements of neighboring Fe-O$_6$ octahedra, the Fe-O bond lengths 
were varied by small amounts ($\approx$ 0.02 - 0.06 $\AA$) within the harmonic oscillation limit around the equilibrium bond lengths. 
The obtained energy versus bond length variation for the three cases for both the compounds are shown in Figs. 4(b) and (c). A parabolic fit of 
the data points provides the estimates of the spin-dependent elastic interactions. 
DMAFeF is found to be weakly ferroelastic with $e_{+-} \simeq \sqrt{e_{++}\times e_{--}}$ while HAFeF is found to be 
strongly ferroelastic with  $e_{+-}> \sqrt{e_{++}\times e_{--}}$, having effective elastic constant of 3.52 K for DMAFeF 
compared to a substantially larger effective elastic constant of 8.93K for HAFeF.

We next turn on to the magnetic  super-exchange  coupling between neighboring Fe(II) centers in HS state. To estimate their values
we calculated the total energies of ferromagnetic and antiferromagnetic  Fe$^{2+}$ spin  configurations, and mapped on to the 
spin Hamiltonian $$\mathcal{H}_{magnetic}=J\times S_i\times S_{j}$$, where $J$ is the magnetic exchange between nearest neighbor 
Fe$^{2+}$ spins, S$_i$ and S$_j$. The difference of the ferromagnetic and antiferromagnetic energies provides the estimate
of $J$, which turned out to be of antiferromagnetic for the compounds, and of values 3.19K for DMAFeF and 2.85K for HAFeF. 
This leads us to conclude that while the magnetic exchanges in the two compounds are of same sign and of similar strengths, the 
spin-dependent elastic interactions are of ferroelastic nature with significantly larger strength for HAFeF compared to DMAFeF. 
Following the previous literature\cite{Banerjee} we also conclude that primary responsible factor in driving cooperativity in these 
formate frameworks is the spin-dependent lattice effect, helped by the magnetic exchange. This is different from the case of 
coordination polymer compounds in which it was found to be entirely driven by the magnetic super-exchanges.\cite{Banerjee,roser} 
The change of amine cation, leads to the change in 
cross-linking hydrogen bonding, and thus to the rigidity of the lattices. Interestingly, the change of amine molecule is also found to
effect the spin-phonon coupling, thereby producing a profound effect on the cooperativity.

\section{Conclusions}

In summary, we show that apart from exhibiting interesting multiferroic properties transition metal formate based hybrid perovskites 
are also potential candidates for exhibiting spin-switching upon application of external stimuli. We demonstrated this through
rigorous first-principles calculations, considering two formate based hybrid perovskite compounds, DMAFeF and HAFeF under 
hydrostatis pressure. We found that dense framework structures of these compound help in building up cooperativity in 
spin-switching, making the phenomena a spin-state transition with appreciable hysteresis effect.
The spin-switching is reflected in associated changes in electronic, magnetic and also possible changes in optical properties. 
This opens up several novel potential applications of these materials, for example, as pressure sensors, as active elements of various types of 
displays, and in information storage and retrieval - an aspect which has remained unexplored so far. Our 
computed values of pressure needed to drive the spin-state transition is found to be in the range of about 2-6 GPa, which should be 
readily achievable in a laboratory set-up of a low to medium pressure diamond anvil cell (DAC). The appreciable hysteresis effect of 2-5 GPa 
associated with these spin-state transitions would make them functional in memory devices for a reasonably wide 
range of pressure. Interestingly, taking the advantage of flexibility of these MOF perovskites to undergo substantial change 
in mechanical properties upon tuning of hydrogen bonds, both the pressure required from the transition, as well as the hysteresis-width 
are found to be tunable by choice of appropriate amine cation. Investigation of microscopics shows elastic properties are vastly 
different between the two studied compounds, lending support to our observation.
 
While in our study, we considered only a particular type of external perturbation, namely the hydrostatic pressure, in principle such SCO
in hybrid perovskites may occur upon light irradiation, or by application of magnetic field, as observed in many of the metal-organic 
complexes.\cite{review} Light induced spin-switching, if can be achieved in hybrid perovskites, will make these materials useful in
application as optical switches as well as for studying phenomena as light-induced excited spin-state trapping. We hope that our study
will stimulate further research, both experimental and theoretical, in discovering this new aspect of hybrid perovskites.

\section{Supporting Information}
The computed structures of DMAFeF and HAFeF, provided as .cif files.

\begin{acknowledgement}
H.B and T.S.D acknowledge the computational support of Thematic Unit of Excellence on Computational Materials Science, funded by Nano-mission of Department of Science and Technology. 
\end{acknowledgement}

\footnotesize{

\bibliography{achemso-demo}

\end{document}